\documentclass[conference]{IEEEtran}
\IEEEoverridecommandlockouts
\usepackage{cite}
\usepackage{amsmath,amssymb,amsfonts}
\usepackage{algorithmic}
\usepackage{graphicx}
\usepackage{textcomp}
\usepackage{xcolor}
\def\BibTeX{{\rm B\kern-.05em{\sc i\kern-.025em b}\kern-.08em
    T\kern-.1667em\lower.7ex\hbox{E}\kern-.125emX}}

\usepackage{xspace}
\usepackage{tikz}
\newcommand{\Design}[0]{ReCIM-HD\xspace}

\begin{document}

\title{ReCIM-HD: A 40~nm On-Device Learning Accelerator with Hyperdimensional Computing via Importance-Based Hybrid Digital–ReRAM CIM \vspace{-10pt}}

\author{\normalsize Keming Fan$^\dagger$$^1$, Chang Eun Song$^\dagger$$^1$, Soumil Jain$^1$, Weihong Xu$^2$, Gopabandhu Hota$^1$, Haichao Yang$^1$,\
Leo Liu$^3$, \\ \normalsize Meng-Fan Chang$^3$, Carlos H. Diaz$^3$,  Gert Cauwenberghs$^1$, Tajana Rosing$^1$, and Mingu Kang$^1$ ($^\dagger$co-first)\\
\normalsize $^1$University of California, San Diego, USA\
\normalsize $^2$EPFL, Switzerland \
\normalsize $^3$Taiwan Semiconductor Manufacturing Company (TSMC)\vspace{-4mm}}

\maketitle

\begin{abstract}
\Design is an on-device learning (ODL) accelerator that integrates gradient-free hyperdimensional computing (HDC) with resistive RAM (ReRAM)-based analog compute-in-memory (CIM).
\Design introduces three key techniques: (1) ultra-low-power HDC-based edge ODL without costly back-propagation operations; (2) non-volatile ReRAM-based storage for class hypervectors and CIM similarity search, enabling energy-efficient inference and power-gating during idle periods without data loss; and (3) a hybrid digital--analog CIM inference flow that protects a small accuracy-critical subset digitally to preserve accuracy.
Additionally, it employs a Kronecker HD encoder (KE) to reduce encoding overhead and a customized instruction set architecture (ISA) to support flexible programmability. Fabricated in 40 nm, \Design achieves 44.17 TOPS/W (hybrid mode) and 75.8 TOPS/W (ReRAM CIM mode), delivering up to 20.1$\times$ higher energy efficiency than the state-of-the-art accelerator.
\end{abstract}


\section{Introduction}
\begin{figure}
    \centering
    \includegraphics[width=\columnwidth]{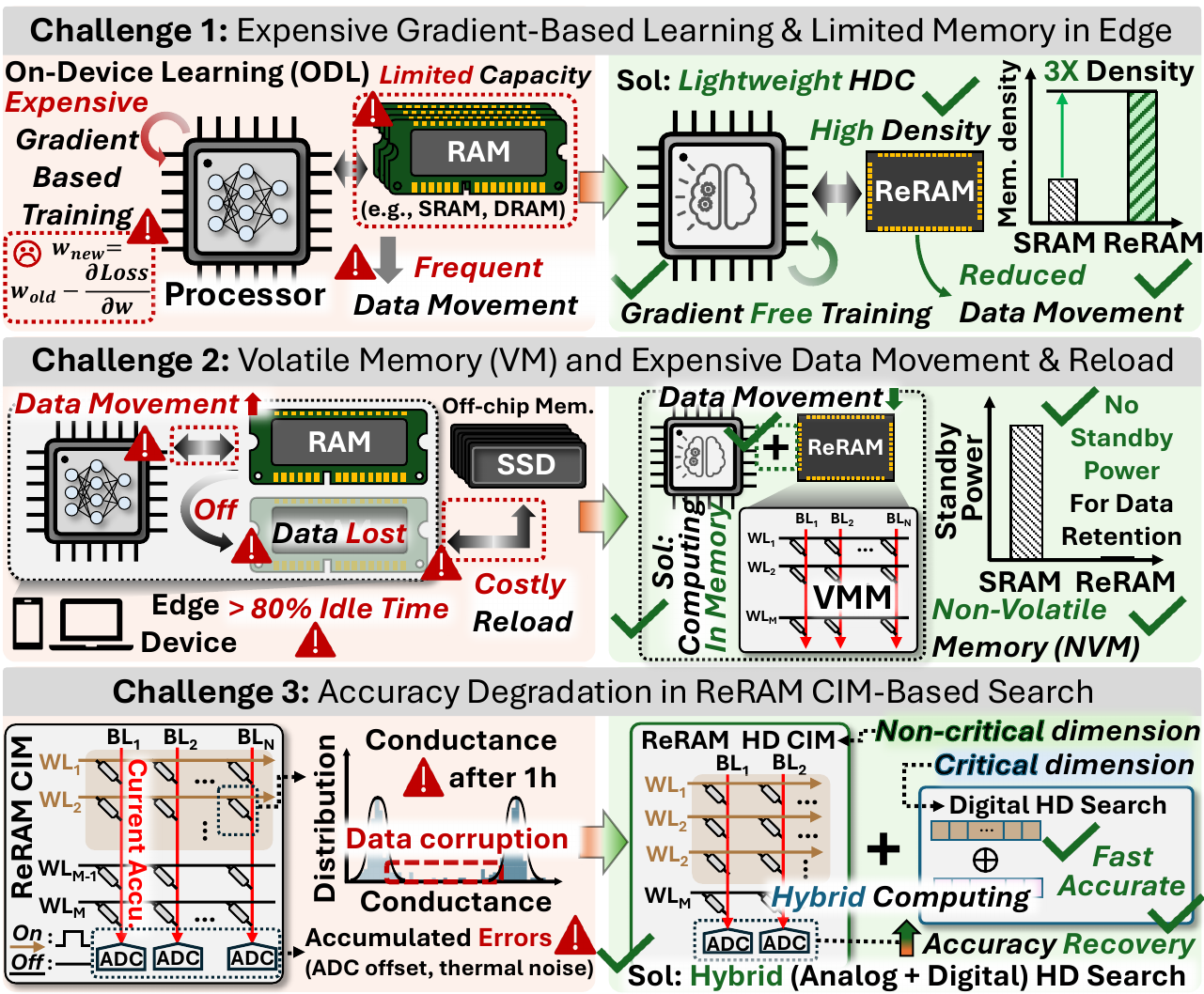}
    \vspace{-20pt}
    \caption{Challenges of ODL at the edge and proposed solutions with hybrid HDC using ReRAM-based CIM.}
    \vspace{-8pt}
    \label{fig: challenge}
\end{figure}

\begin{figure}
    \centering
    \includegraphics[width=\columnwidth]{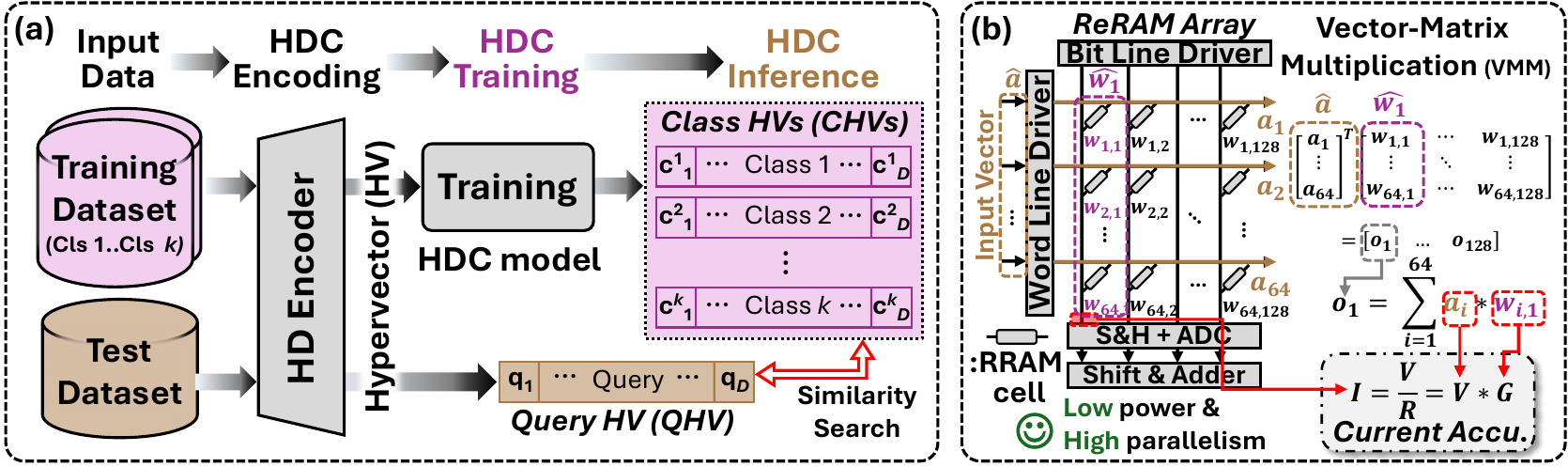}
    \vspace{-20pt}
    \caption{Overview of (a) HDC and (b) ReRAM CIM.}
    \vspace{-15pt}
    \label{fig: hdc}
\end{figure}

On-device learning (ODL) is crucial for continual adaptation to environmental changes while preserving data privacy and security. However, edge platforms operate under severe constraints in compute capability, memory capacity, and data bandwidth~\cite{r1_karunaratne2022memory, r2_ma2023cost}.
Existing ODL accelerators~\cite{r3_den2026chameleon, r4_song2025clo, r5_yang2024fsl, r6_huang2023nonvolatile, r7_li2021sapiens} face three key challenges as described in Fig.~\ref{fig: challenge}.
\textbf{(C1)} Gradient-based training incurs a high computational cost, and limited on-chip memory forces frequent data movement between processors and memory.
\textbf{(C2)} Volatile memory (e.g., DRAM and SRAM) requires frequent model reloads when edge devices, which are largely idle due to intermittent inputs, are power-gated~\cite{r4_song2025clo}.
While non-volatile memory (NVM)-based compute-in-memory (CIM) (Fig.~\ref{fig: hdc}(b)) alleviates reload overhead~\cite{r8_kazi2014energy}, \textbf{(C3)} analog CIM and device non-idealities result in significant accuracy degradation.

To tackle these challenges, \Design introduces several key techniques. \textbf{(S1)} We adopt gradient-free hyperdimensional computing (HDC) learning for efficient ODL (Fig.~\ref{fig: hdc}(a)), using simple and highly parallelizable bit-wise operations. However, conventional HDC relies on high-dimensional hypervectors (HVs) stored in large associative memories, incurring substantial storage and data movement overhead. We therefore exploit high-density resistive RAM (ReRAM) as an NVM substrate to store HVs in place.
\textbf{(S2)} This enables near-zero standby power during idle periods under power-gating while preserving stored data, and low-power, highly parallel CIM-based similarity search to reduce data movement. \textbf{(S3)} To mitigate non-idealities from analog CIM and device imperfection, \Design employs an importance-based hybrid digital–analog execution scheme, where only a small accuracy-critical portion of dimensions is processed digitally while the majority is mapped to ReRAM CIM. Furthermore, \Design incorporates a Kronecker HD encoder (KE) to reduce encoding overhead and a customized instruction set architecture (ISA) to flexibly program both HDC and CIM operations.

\begin{figure}
    \centering
    \includegraphics[width=1\columnwidth]{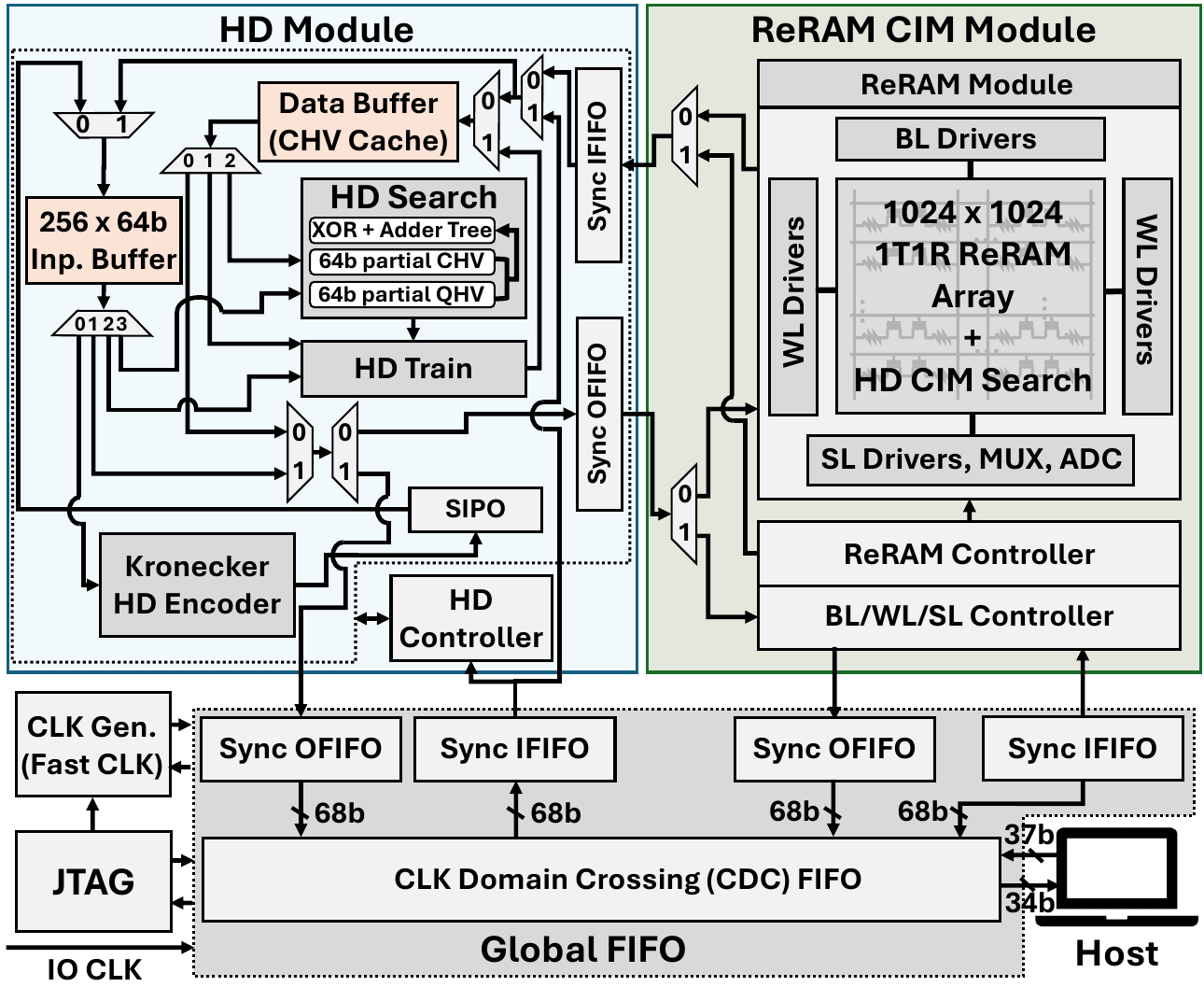}
    \caption{Proposed \Design hardware with key components including HD module, ReRAM CIM module, and global FIFO.}
    \label{fig: overview}
    \vspace{-10pt}
\end{figure}

\section{Proposed Design} Fig.~\ref{fig: overview} shows \Design overall architecture.
The HD module supports the full HDC pipeline (encoding, training, and similarity search) digitally,
while the ReRAM CIM module stores bipolar (+1/-1) class HVs (CHVs) and performs inference via CIM-based analog similarity search, with a FIFO interface managing data exchange and synchronization.

\begin{figure}[h]
    \centering
    \includegraphics[width=\columnwidth]{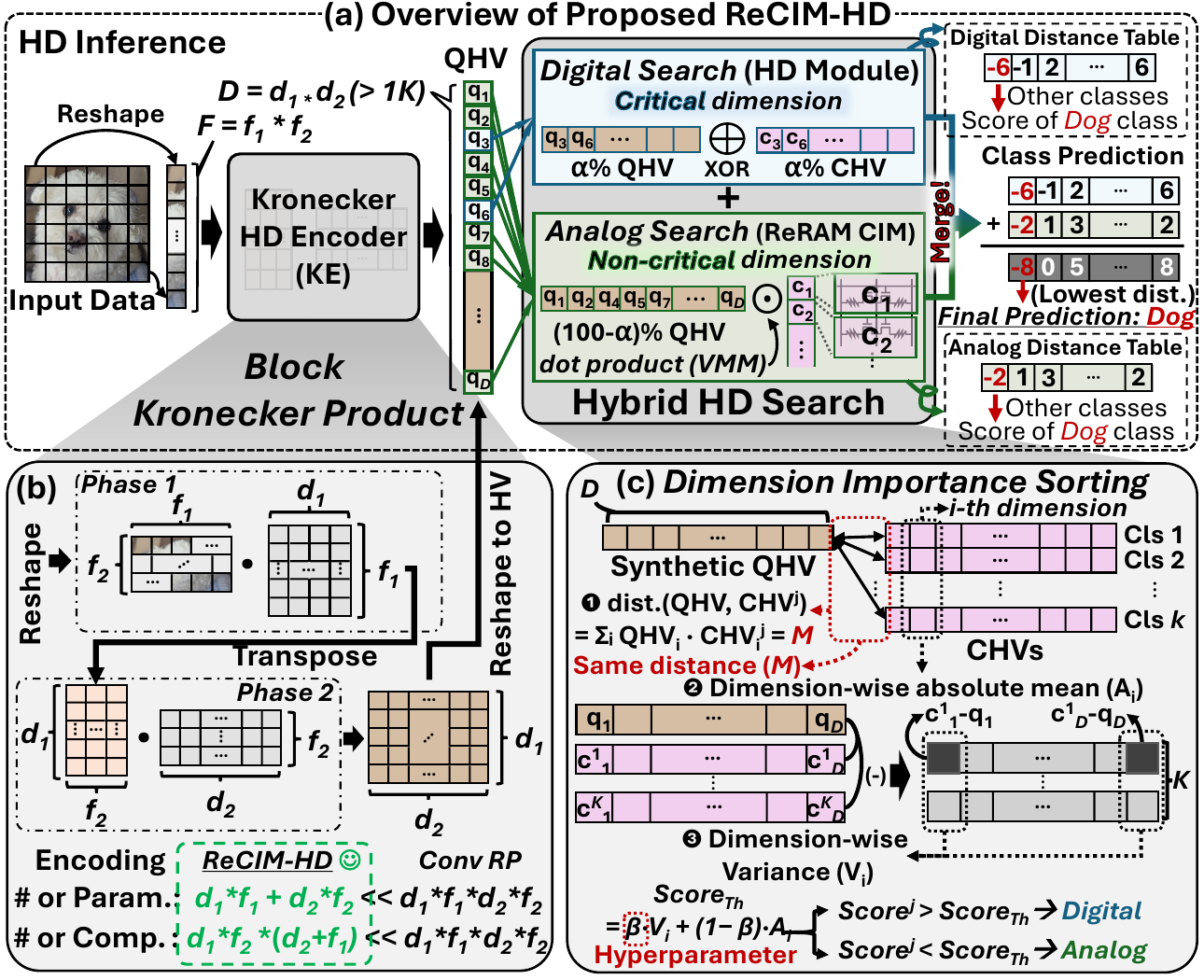}
    \vspace{-20pt}
    \caption{(a) Proposed \Design dataflow \& HD inference via hybrid HD search, (b) overview of KE, and (c) identification of critical dimensions based on importance sorting.}
    \vspace{-20pt}
    \label{fig: encoder}
\end{figure}

\begin{figure}[h]
    \centering
    \includegraphics[width=1\columnwidth]{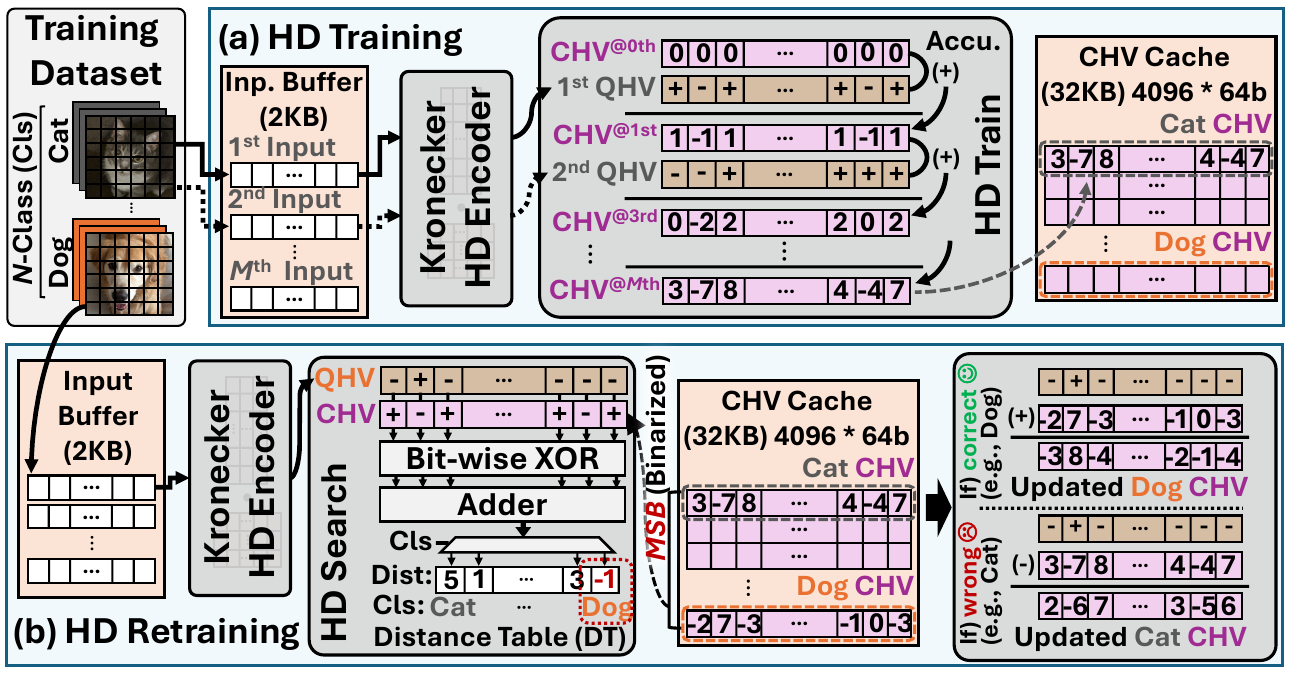}
    \vspace{-15pt}
    \caption{HDC operations of (a) training and (b) retraining.}     \vspace{-15pt}
    \label{fig: details}
\end{figure}

\subsection{\Design Inference and Training Dataflow}

Fig.~\ref{fig: encoder}(a) shows the proposed \Design data flow during HD inference. Input features are encoded into a bipolar (+1/-1) query HV (QHV) by the KE~\cite{r10_xu2025hypermetric} (Fig.~\ref{fig: encoder}(b)), which consists of two phases, each performing a reshape followed by matrix multiplication to generate the encoded HV.
The KE reduces storage by $204.8\times$ and computation by $21.3\times$ compared to random projection-based encoding~\cite{r11_hernandez2021reghd}
by factorizing a large projection matrix into two smaller matrices.
Then, the encoded QHV is compared against all stored bipolar CHVs through dot-product–based similarity search to generate a distance table where the class with the smallest distance is taken as the final prediction.
To balance efficiency and accuracy, \Design proposes a \textit{hybrid digital--analog HD search} scheme in which accuracy-critical dimensions, identified one-time offline, are processed digitally through the HD module while the remaining majority are mapped to ReRAM and processed via analog CIM (Section~\ref{sec: RRAM}).
As shown in Fig. 4(c), we construct a synthetic vector that yields identical similarity scores across all CHVs, and analyze each dimension by measuring the mean ($A_i$) and variance ($V_i$) of the deviations of CHVs from this synthetic reference for the $i$-th dimension.
Based on $\beta \cdot V_i + (1-\beta) \cdot A_i$, where $\beta$ is a hyperparameter, all dimensions are ranked, and the top $\alpha$ fraction of dimensions is selected as the critical set, which is static during inference.
For the digital portion, the critical dimensions of each CHV are stored in the on-chip cache.
In each cycle, 64 MSBs of a CHV segment are fetched and compared against the corresponding QHV segment using the XOR and adder tree to compute partial similarity scores.
For the analog portion, the remaining dimensions ($\approx$ 90\%, i.e., $1-\alpha$) are stored in ReRAM and evaluated through analog CIM.
The two similarity scores are summed to produce the final score in the distance table for classification.

Fig.~\ref{fig: details} shows the details of on-device HD training and retraining.
(a) During HD training, the HD module encodes each sample into a bipolar QHV using the KE and accumulates same-class QHVs to form CHVs. (b) HD retraining further updates the initial CHVs by first performing HD inference.
If the predicted class is correct, the corresponding CHV is reinforced by adding the QHV; otherwise, the QHV is subtracted from the misclassified CHV to refine class representations.

\vspace{-10pt}

\begin{figure}[h]
    \centering
    \includegraphics[width=0.97\columnwidth]{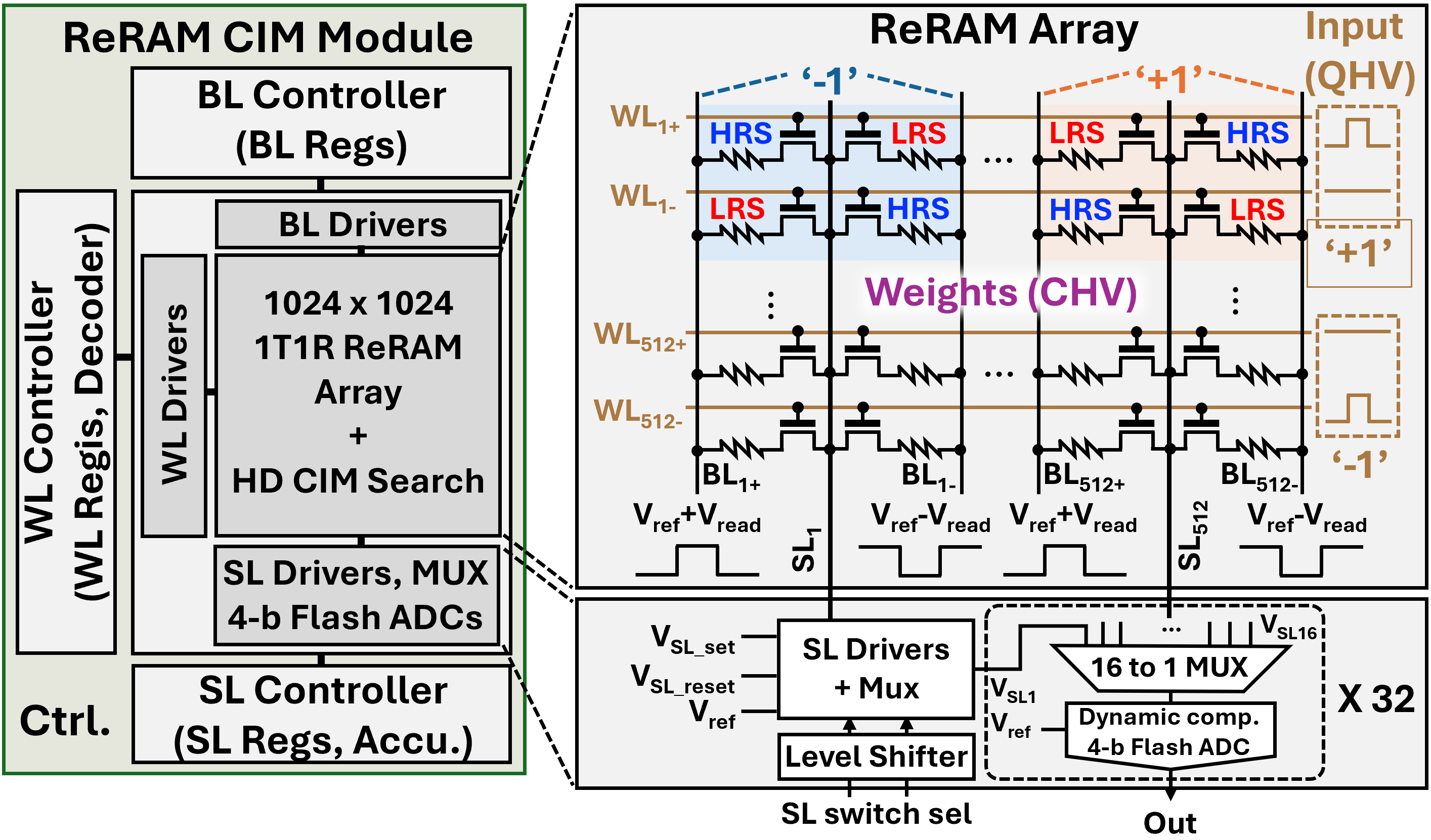}
    \vspace{-10pt}
    \caption{Overview of ReRAM CIM module and its CIM operation
for HDC-based vector similarity calculation.}
\vspace{-10pt}
    \label{fig: RRAM}
\end{figure}

\subsection{ReRAM Macro} \label{sec: RRAM}
\begin{figure}[h]
    \centering
    \includegraphics[width=\columnwidth]{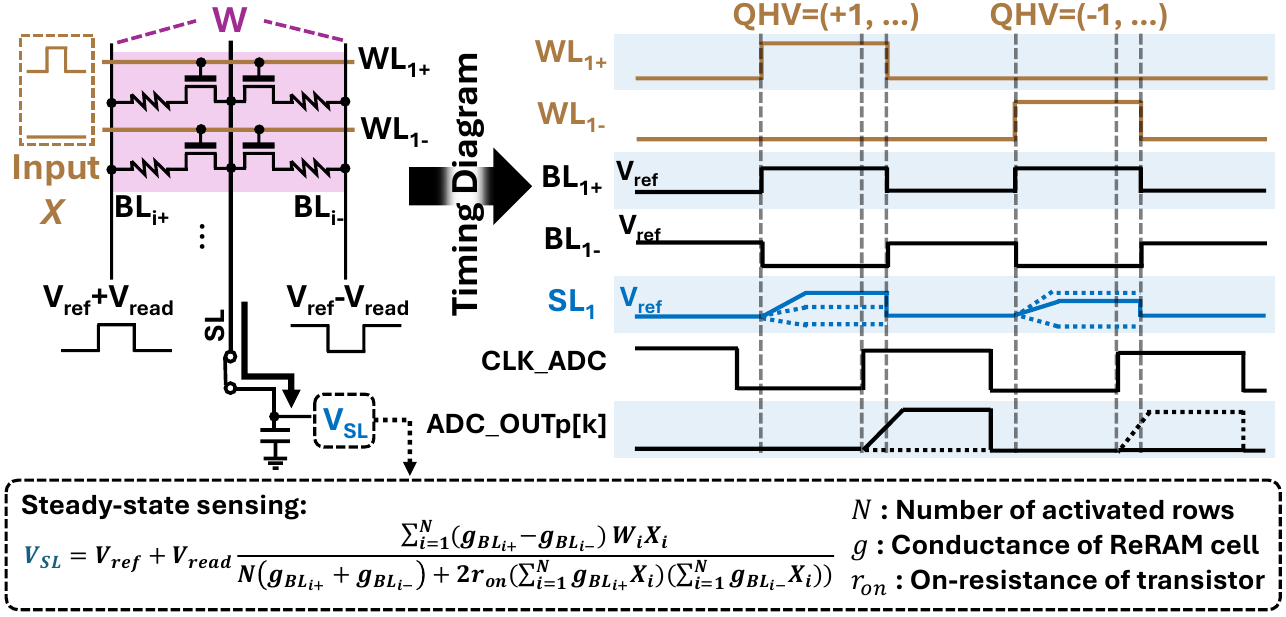}
    \vspace{-22pt}
    \caption{ReRAM analog CIM operation, and timing diagram.}
    \label{fig: RRAM_2}
    \vspace{-15pt}
\end{figure}

Fig.~\ref{fig: RRAM} outlines the ReRAM module, which consists of a digital ReRAM controller, a 1k$\times$1k 1T1R ReRAM array, and its peripherals, including drivers and ADCs.
The ReRAM array supports two operating modes.
For normal read/write operations, \Design uses a single-differential mapping with two cells per bipolar bit (+1/-1) to maximize storage density.
For CIM-based similarity search, it adopts a double-differential weight mapping scheme using four cells per bipolar bit (i.e., representing +1/-1 as shown in Fig.~\ref{fig: RRAM}), improving accuracy and noise tolerance.
During CIM, each CHV is stored per column, and if the vector dimension exceeds the number of rows, the CHV is split into segments and distributed over multiple columns.
The bipolar input vector QHV is encoded through wordlines (WLs), while BL$+$ and BL$-$ are biased at $V_{\mathrm{ref}}+V_{\mathrm{read}}$ and $V_{\mathrm{ref}}-V_{\mathrm{read}}$, respectively, independent of the input.
Depending on the stored CHV polarity and the applied QHV polarity, each activated cell contributes a positive or negative current to the shared sourceline (SL). At steady state, the accumulated voltage on the SL capacitor ($V_\textrm{SL}$) represents the dot-product-based similarity score, as shown in Fig.~\ref{fig: RRAM_2}.
The resulting SL voltage is then multiplexed and converted by a 4-bit flash analog-to-digital converter (ADC).
Each ADC is shared across 16 SLs to save area while matching throughput.
These partial sums are further accumulated in the ReRAM controller to cover full dimensions, and are then forwarded to HD module for end-to-end inference.

\subsection{Customized ISA} 
\begin{figure}
    \centering
    \includegraphics[width=0.97\columnwidth]{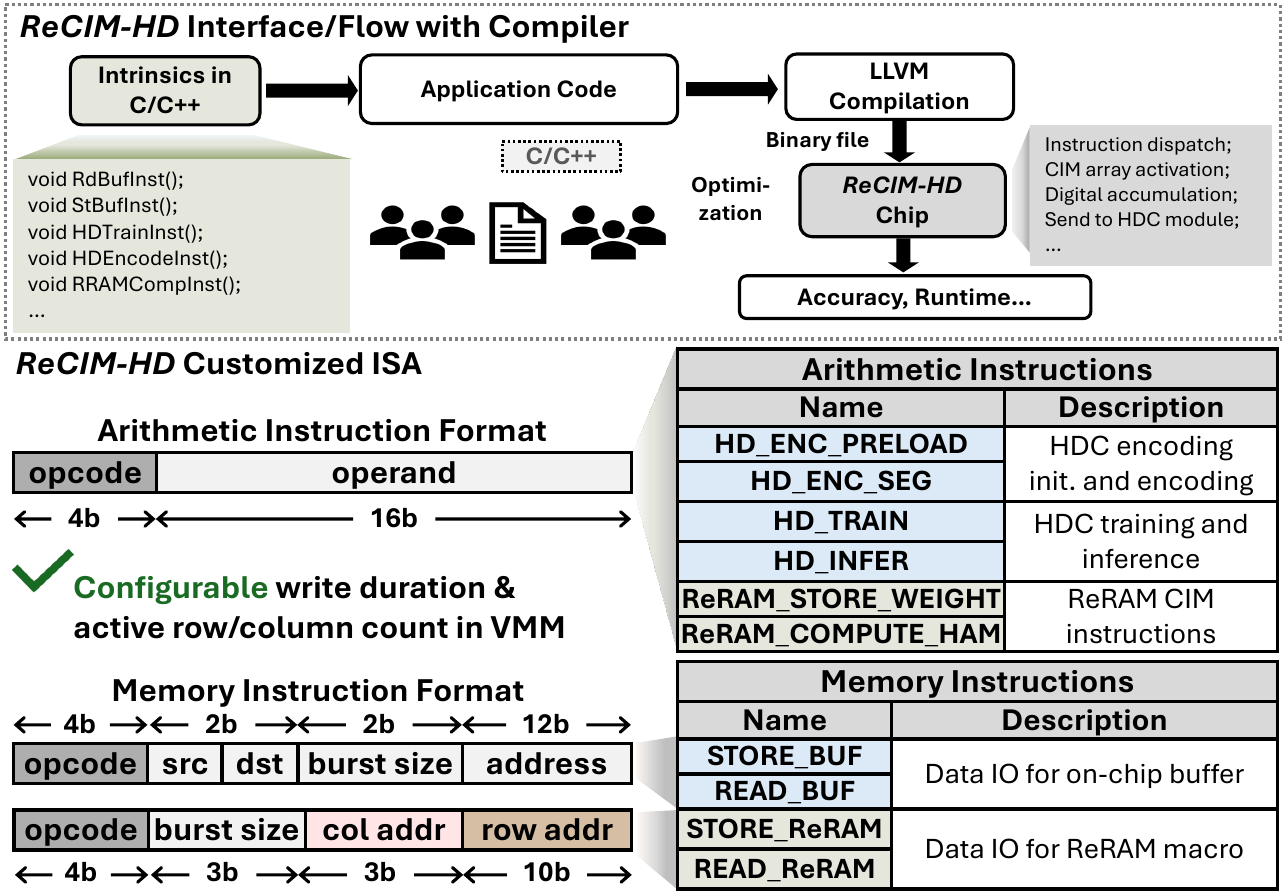}
    \vspace{-10pt}
    \caption{Customized ISA and compiler interface for fast and flexible programming for \Design.}
    \vspace{-10pt}
    \label{fig: ISA}
\end{figure}

Fig.~\ref{fig: ISA} presents the customized ISA of \Design and its software-to-hardware execution flow.
The ISA adopts a 20-bit instruction format consisting of a 4-bit opcode and a 16-bit operand, enabling flexible programmability across both digital HDC operations and ReRAM CIM execution.
\Design supports basic HDC arithmetic instructions, such as encoding and training, as well as memory-related instructions, such as store and read.
For CIM execution, the ReRAM controller exposes several programmable knobs through the ISA. It supports configurable write durations to balance write energy and latency against programming accuracy.
Configurable row-level parallelism is available during vector–matrix multiplication (VMM), with 16/32/64-row activation modes, along with configurable column parallelism through burst-size–controlled column activation.
We implement C/C++ intrinsics for each instruction so that inline assembly in high-level code compiles directly to the corresponding bytecode. This lightweight ISA/compiler stack enables flexible programmability while preserving low-overhead hardware control.

\section{Silicon Measurement and Results}
\vspace{-5pt}
\begin{figure}[ht]
    \centering
    \includegraphics[width=\columnwidth]{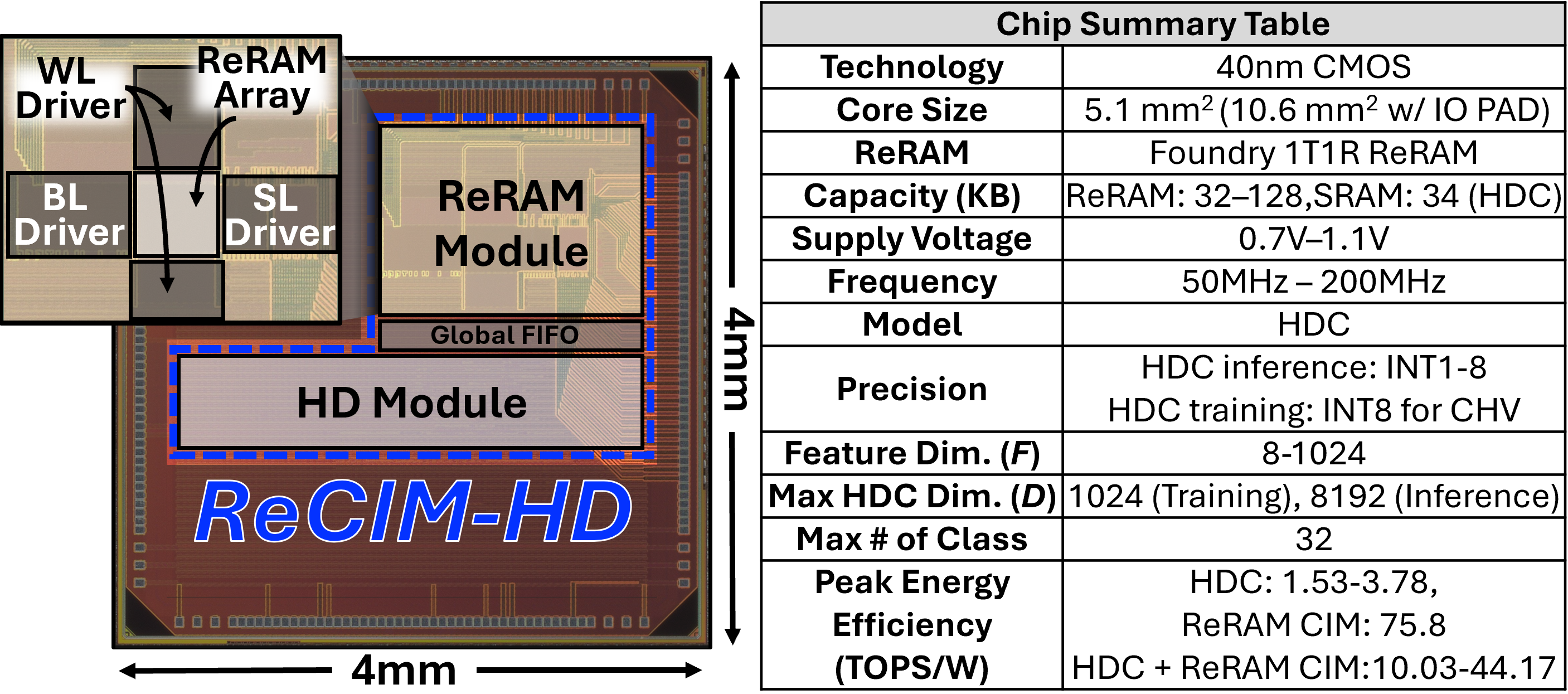}
    \vspace{-22pt}
    \caption{ReCIM-HD die photo with chip specifications.}
    \label{fig: die}
    \vspace{-10pt}
\end{figure}
\begin{figure}[h]
    \centering
    \includegraphics[width=\columnwidth]{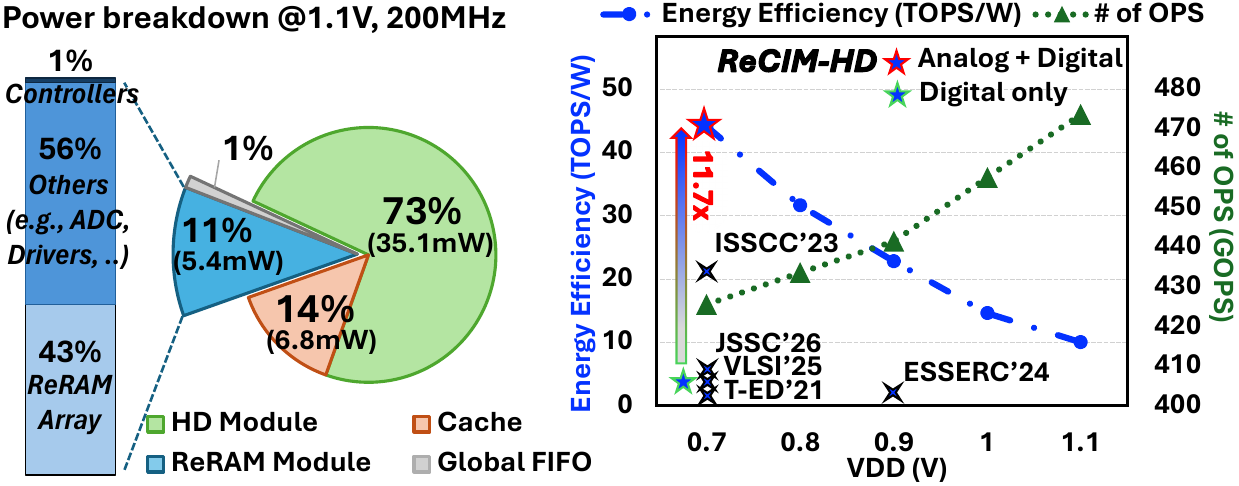}
    \vspace{-20pt}
    \caption{\Design power breakdown and energy efficiency of ReCIM-HD across different supply voltages of the HD module.}
    \vspace{-10pt}
    \label{fig: power}
\end{figure}
\begin{figure}[t]
    \centering
    \includegraphics[width=\columnwidth]{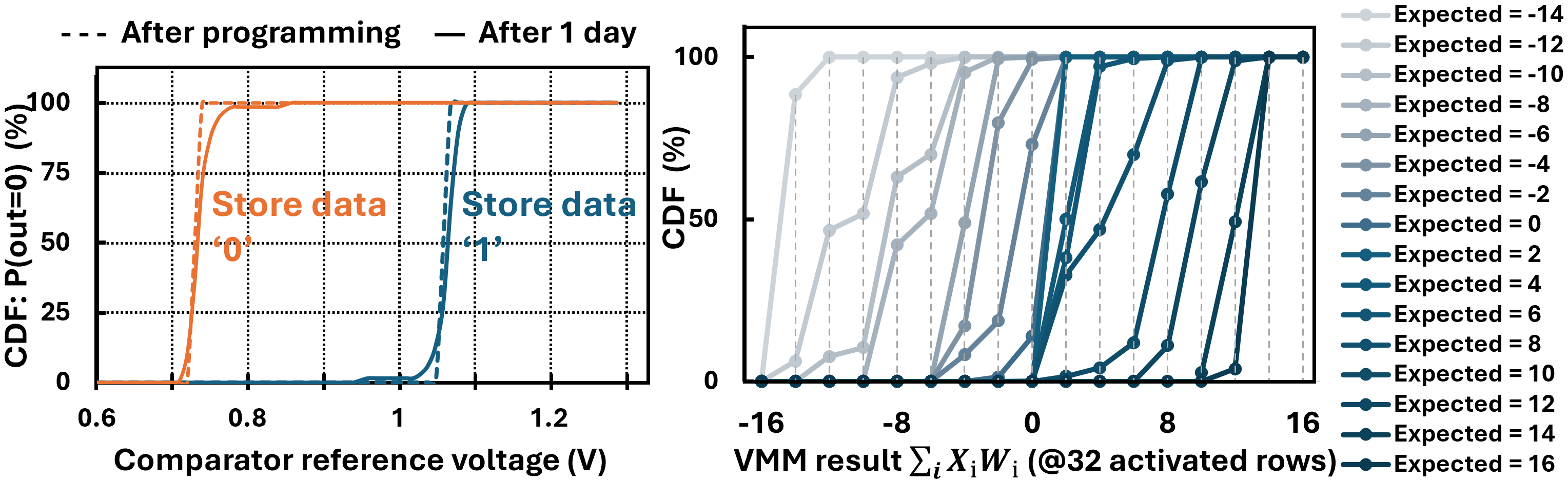}
    \vspace{-20pt}
    \caption{Measured results of ReRAM read operation and VMM CIM. }
    \vspace{-10pt}
    \label{fig: measured_acc}
\end{figure}

Fig.~\ref{fig: die} shows the chip micrograph and specifications of \Design, fabricated in 40~nm CMOS with a die area of 5.1~mm$^2$, integrated with a foundry 128~KB 1T1R ReRAM macro. The HD module and ReRAM CIM module can operate over a frequency range of 50--200~MHz.
The power breakdown in Fig.~\ref{fig: power} shows that the ReRAM module consumes only 5.4~mW, accounting for 11\% of the total chip power while achieving 75.8~TOPS/W,  highlighting the high energy efficiency of analog CIM.
By using hybrid search, \Design achieves up to 44.17~TOPS/W during inference at 0.7--1.1~V  when the critical dimension ratio is $\alpha=10\%$.
We measure both ReRAM read operations for storage and CIM operations for in-memory vector-matrix multiplication (VMM). Each ReRAM cell’s resistance is characterized by collecting the cumulative distribution function (CDF) of readout values from 1024 cells using the on-chip 4-bit flash ADC configured as a 1-bit comparator while sweeping its reference voltage. Fig.~\ref{fig: measured_acc} shows the binary readout distributions measured immediately after programming and again one day later.
It also shows the CIM results as  CDF of expected and measured VMM outputs with 32 activated rows per cycle during inference.

\begin{figure}[t]
    \centering
    \includegraphics[width=\columnwidth]{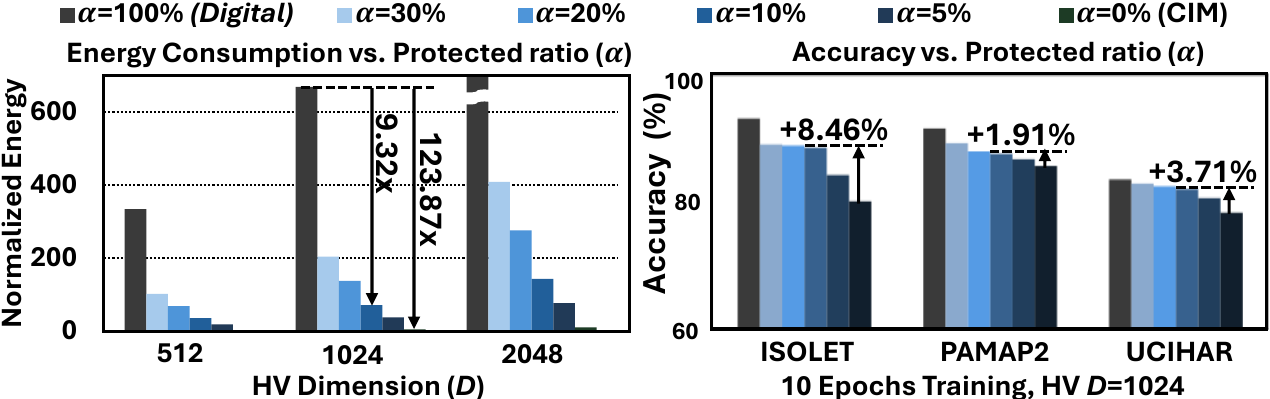}
    \vspace{-20pt}
    \caption{Energy consumption and accuracy of \Design across different protected ratios ($\alpha$). }
    \vspace{-15pt}
    \label{fig: acc}
\end{figure}

End-to-end system accuracy is evaluated on three benchmarks, including ISOLET for speech recognition~\cite{isolet_54},
PAMAP2~\cite{r12_pamap2} for physical activity monitoring,
and UCIHAR~\cite{r14_human_activity_recognition_using_smartphones_240} for human activity recognition.
Fig.~\ref{fig: acc} summarizes the accuracy and energy efficiency across different critical dimension ratios $\alpha$.
While a fully analog CIM implementation achieves 123.87$\times$ higher energy efficiency than fully digital processing, this gain comes at the cost of reduced accuracy.
By contrast, the hybrid HD search in \Design effectively balances accuracy and efficiency: with only $\alpha=10\%$, it improves accuracy by 1.91--8.46\% over the CIM-only design while still reducing energy consumption by 9.32$\times$ relative to the digital-only implementation. By increasing $\alpha$, the accuracy can be further recovered, at the cost of higher energy consumption.
 \vspace{-8pt}

\begin{table}[h]
    \centering
    \caption{Comparison with state-of-the-art models.}
    \vspace{-5pt}
    \includegraphics[width=0.95\columnwidth]{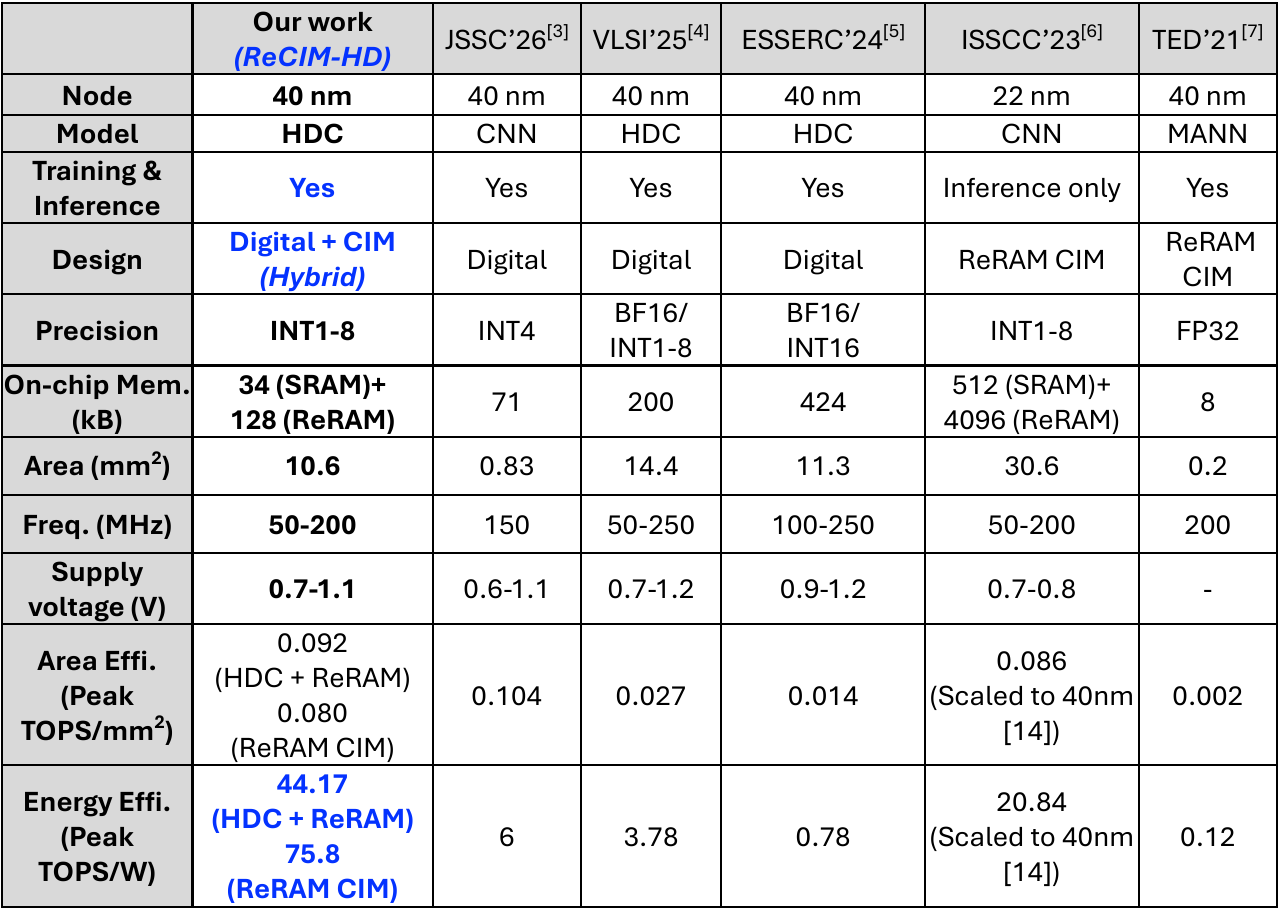}
    \label{tab: comparison}
\end{table}

Table~\ref{tab: comparison} compares \Design with prior accelerators.
Unlike prior ReRAM CIM designs~\cite{r6_huang2023nonvolatile} focusing on inference only, ReCIM-HD supports both learning and inference.
CNN-based accelerators enable training~\cite{r3_den2026chameleon} but rely on costly gradient operations, while HDC accelerators~\cite{r4_song2025clo, r5_yang2024fsl} allow lightweight learning but are fully digital and do not utilize the benefits of analog CIM.
\Design is the first silicon prototype to combine HDC-based ODL with CIM-based inference, achieving 20.1--56.6$\times$ and 3.4--6.6$\times$ higher energy and area efficiency, respectively, than prior works by leveraging the high efficiency of analog CIM and the high density of ReRAM.

\vspace{-3pt}

\section{Conclusion}
This paper presents \Design, the first silicon prototype to combine HDC-based on-device learning with ReRAM CIM-based inference.
\Design enables low-power gradient-free edge learning, non-volatile model retention under power-gating, and energy-efficient in-memory inference.
To mitigate analog-CIM non-idealities, \Design further adopts an importance-based hybrid digital--analog search that protects only a small accuracy-critical subset of dimensions digitally.
Finally, it employs an optimized HD encoder and a customized ISA.
Fabricated in 40 nm CMOS with a foundry 128~KB 1T1R ReRAM macro, \Design achieves up to 20.1$\times$ higher energy efficiency than SOTA accelerators~\cite{r4_song2025clo}.

\vspace{-3pt}

\section*{Acknowledgment}
This work was supported by TSMC and in part by PRISM and CoCoSys, centers in JUMP 2.0, an SRC program sponsored by DARPA.

\vspace{-10pt}

\nocite{r15_stillmaker2017scaling}

\bibliographystyle{IEEEtran}
\bibliography{reference.bib}

\end{document}